\documentclass[%
preprintnumbers,
nofootinbib,
 amsmath,amssymb,
 aps,
 showkeys,
]{revtex4-2}

\usepackage{graphicx}
\usepackage{dcolumn}
\usepackage{bm}

\begin{document}

\preprint{KEK-TH-2775, J-PARC-TH-0327}

\title{On the role of induced electric field in the time-dependent Aharonov-Bohm effect}


\author{Masashi Wakamatsu}
 \email{wakamatu@post.kek.jp}
\affiliation{%
KEK Theory Center, Institute of Particle and Nuclear Studies,
High Energy Accelerator Research Organization (KEK),
Oho 1-1, Tsukuba, 305-0801, Ibaraki, Japan
}%




\date{\today}

\begin{abstract}
Whether the time-dependent Aharonov-Bohm (AB) effect even exists or not 
has been the subject of long-standing debate. 
There are two factors complicating the problem.
First, in the closed spacetime line integral of the vector potential that is thought 
to give the AB-phase shift, how to treat the time-varying vector potential
is highly nontrivial.
Second, the time-varying magnetic flux generates induced electric field even
outside the solenoid. In the present paper, motivated by a recent work 
by Gao, we re-investigate the role of the induced electric field with the utmost
care. This analysis reveals a highly nontrivial effect of the induced electric field, 
which turns out to be useful for verifying the very existence of the 
time-dependent AB-effect.
\end{abstract}




\maketitle


\section{Introduction}
\label{Section:s1}

The importance of the Aharonov-Bohm (AB) effect in the contemporary physics 
can never be overstated \cite{ES1949,AB1959}. 
Although its evidence was eventually confirmed by ingenious experiments 
by Tonomura et al. \cite{Tonomura1986, Osakabe1986}, 
a lot of debates on the theoretical interpretation
of this phenomenon is continuing even now. (For review, see \cite{Peshkin1981,
OP1985, PT1989, WKZZ2018}, for example.)
There is another even more controversial problem called the time-dependent
AB-effect, which deals with the circumstances in which the magnetic field inside the
solenoid varies time-dependently.
Theoretically complex nature of the problem has prevented us from 
reaching a confident conclusion on whether such a phenomenon 
even exist or not. 

\vspace{1mm}
In a previous paper \cite{Waka2025}, we tried to resolve one long-standing issue over 
the time-dependent Aharonov-Bohm effect.   
The question was the validity or failure of the remarkable claim by Singleton 
and his collaborators, who insisted that the time-dependent part of the AB-phase shift due 
to the magnetic vector potential is precisely cancelled by the effect of 
the induced electric field arising from the time-varying magnetic flux  inside
the solenoid so that only the AB-phase shift corresponding to the time-independent
part of the magnetic field remains \cite{SV2013,MS2014,AS2016}. 
Through a careful analysis of the 4-dimensional Stokes theorem in the
Minkowski space, we showed that the failure of their cancellation argument comes 
from an unjustified treatment of area-integral side of the 4-dimensional Stokes theorem. 
Its correct treatment led us to the conclusion that the time-dependent Aharonov-Bohm 
effect is most likely to exist. (We recall that most authors in the past papers favors 
the existence of the time-dependent AB-effect, although their theoretical treatments 
as well as predictions are fairly divergent \cite{LYGC1992, ADC2000, GNS2011, 
JZWLD2017, CM2019}.) 
Besides, we derived a closed-form analytical expression for the time-dependent 
AB-phase shift. Unfortunately, this answer was obtained by restricting to some special 
situations in which the effect of the induced electric field can be considered negligibly
small. It happens in two specific circumstances. One is the case where the time-variation 
of the magnetic flux is slow enough. 
The other is the case where the magnetic field in the solenoid is
sinusoidally and rapidly oscillating and the time period of this oscillation is much
shorter than the time scale between the emission time of the electron beam and 
its arrival  time on the interference screen.

\vspace{1mm}
Remarkably, in a recent paper \cite{Gao2025}, Gao showed that the same answer as 
ours can be obtained even in account of the induced electric field effect.
This is a highly non-trivial finding in view of the fact that the motion of the electron
is evidently thought to be influenced by the Lorentz force due to the induced electric 
field generated by the time-varying magnetic flux of the solenoid. 
This motivates us to reinvestigate the role of the induced electric field on the 
time-dependent AB-effect by utilizing Cao's elementary but elegant 
formulation but with extra care. 
We shall show that this analysis exposes a highly nontrivial effect of the induced
electric field, which was overlooked or neglected in the argument by Gao.

\vspace{1mm}
The paper is organized as follows. In sect.II, we investigate the role of the induced
electric field by basically following the elementary formulation by Gao, but paying
minute attention to its nontrivial function neglected in his paper.
Next, in sect.III, we shall discuss some serious questions which still remain in the 
theoretical treatment explained in the previous section. We also propose a possibly 
simplest measurement which would provide manageable means for verifying the 
very existence of the time-dependent AB-effect.

\section{\label{sec2}The role of induced electric field in the time-dependent
AB-effect}
\label{Section:s2}

We start with the familiar expression for the magnetic field distribution generated
by an infinitely-long solenoid with radius $R$ placed along the $z$-axis, on the 
surface of which uniform but time-dependent electric current is flowing.
This magnetic field distribution can be expressed as
\begin{equation}
 \bm{B} (\bm{x}, t) \ = \ \left\{ \begin{array}{cc}
 B (t) \,\bm{e}_z \ & \ (\rho < R), \\
 \vspace{-3mm} \\
 0 \ & \ (\rho > R). 
 \end{array} \right. 
\end{equation}
Here, we use the cylindrical coordinates $\bm{x} = (\rho, \phi, z)$ with
$\rho = \sqrt{x^2 + y^2}$. By using the total magnetic flux
$\Phi (t) \equiv 2 \,\pi R^2 B (t)$, the vector potential generating the above 
magnetic field is given as
\begin{equation}
 \bm{A} (\bm{x}, t) \ = \ \left\{ \begin{array}{cc}
 \frac{\rho \,\Phi (t)}{2 \,\pi \,R^2} \,\,\bm{e}_\phi \ & \ (\rho < R), \\
 \vspace{-3mm} \\
 \frac{\Phi (t)}{2 \,\pi \,\rho} \,\,\bm{e}_\phi \ & \ (\rho > R),
 \end{array} \right. 
\end{equation}
where $\bm{e}_\phi$ stands for the circumferential unit vector.
Since the net electric charge on the solenoid surface is zero, the
scalar potential $A^0$ can be set zero without loss of generality.
Then, the induced electric field is given by
\begin{equation}
 \bm{E} (\bm{x}, t) \ = \ - \,\frac{\partial}{\partial t} \,\bm{A} (\bm{x}, t) \ = \  
 \left\{ \begin{array}{cc}
 - \,\frac{\rho \,\dot{\Phi} (t)}{2 \,\pi \,R^2} \,\,\bm{e}_\phi \ & \ (\rho < R), \\
 \vspace{-3mm} \\
 - \,\frac{\dot{\Phi} (t)}{2 \,\pi \,\rho} \,\,\bm{e}_\phi \ & \ (\rho > R), \\
 \end{array} \right.
\end{equation}
where $\dot{\Phi} (t)$ represents time derivative of $\Phi (t)$, i.e.
$\dot{\Phi} (t) \equiv \frac{d \Phi (t)}{d t}$.
Undoubtedly, the difficult nature of the time-dependent Aharonov-Bohm effect 
is related to the fact that this induced electric field is nonzero even outside the
shielded solenoid.

\vspace{2mm}
After detailed analysis of the 4-dimensional Stokes theorem in the Minkowskii
space, our previous analysis reconfirmed the fact that the time-dependent AB-phase 
shift is given by the closed space-time line-integral of the 4-vector potential $A_\mu (x)$
as
\begin{equation}
 \phi_{AB} \ = \ - \,e \,\oint_{C_1 - C_2}  \, A_\mu (x) \,d x^\mu \ = \ 
 e \,\left\{\, \int_{C_1} \,\bm{A} (\bm{x}, t) \cdot d \bm{x} \ - \ 
 \int_{C_2} \,\bm{A} (\bm{x}, t) \cdot d \bm{x} \,\right\},
\end{equation}
where the closed space-time path $C_1 - C_2$ is taken so as to surround the solenoid.
(In the above equation, we have used the relation $A^0 = 0$. We also point out that,
throughout the paper, the charge of the electron is taken as $- \,e$ with
$e = |e| > 0$.) 
As emphasized in \cite{Waka2025}, an easy mistake in evaluating this line integral in 
the case of time-dependent magnetic flux comes from the lack of the solid understanding of 
fact that the spatial variable $\bm{x}$ and the time variable $t$ cannot be treated as
independent integration variables. 
For the convenience of readers, we briefly repeat the derivation of the time-dependent
AB-phase shift carried out in \cite{Waka2025}, where the effect of the induced electric 
field is neglected. 

\vspace{2mm}
\begin{figure}[ht]
\begin{center}
\includegraphics[width=13.0cm]{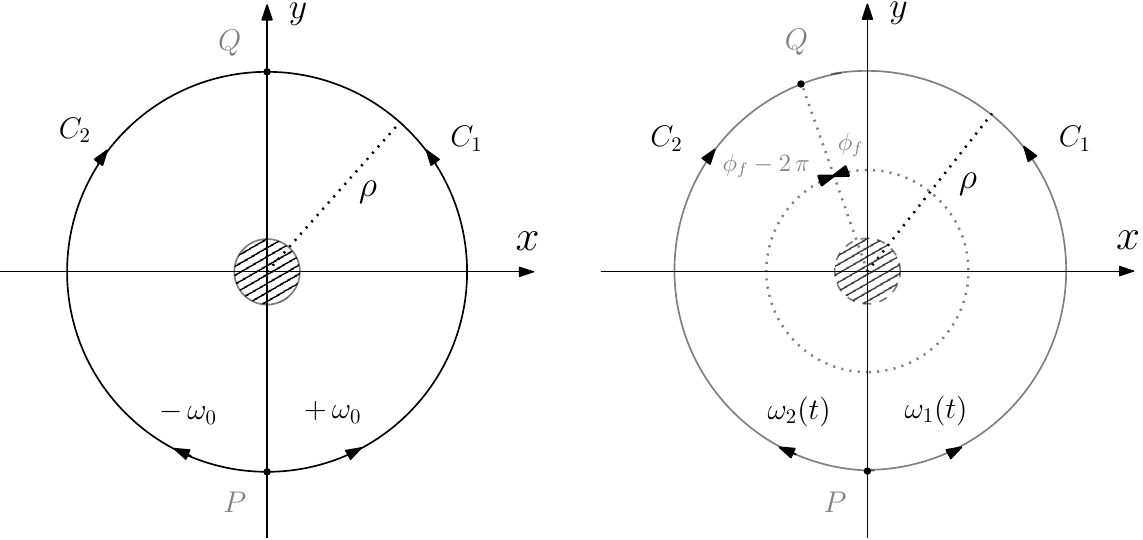}
\caption{Schematic picture to consider the Aharonov-Bohm effect. 
The left panel corresponds to the familiar case of time-independent magnetic flux, 
whereas the right panel corresponds to the case of time-dependent magnetic flux.}
\label{Fig:time-dep.B}
\end{center}
\end{figure}

The left panel of Fig.\ref{Fig:time-dep.B} shows a schematic picture, which provides
us with the theoretical basis of discussing the familiar AB-effect with 
time-independent magnetic flux.
Suppose that, at the initial time $t_i = 0$, the electron beams are emitted
from the point $P$ with the opposite angular velocity $+ \,\omega_0$
and $- \,\omega_0$. They respectively travel along the path 
$C_1$ an $C_2$ and arrive at the point $Q$ at $t = t_f$.
If we neglect the Lorentz force due to the induced electric field, the angular
velocities of the two electrons remain constants in time. 
The path element of $C_1$ is then parametrized as 
$d \bm{x} = \bm{e}_\phi \,\rho \,d \phi = \bm{e}_\phi \,\rho \, \omega_0 \,d t$. 
The line integral of the vector potential along the path $C_1$ can therefore be 
evaluated as follows : 
\begin{equation}
 \int_{C_1} \,\bm{A} (\bm{x}, t) \cdot d \bm{x} \ = \ \int_0^{t_f} \, 
 \frac{\Phi (t)}{2 \,\pi \,\rho} \,\,\rho \,\omega_0 \,d t  \ = \ 
 \frac{\omega_0}{2 \,\pi} \, \int_0^{t_f} \Phi (t) \, d t.
\end{equation}
By using the relation $\omega_0 \,t_f = \phi_f = \pi$, this gives
\begin{equation}
 \int_{C_1} \,\bm{A} (\bm{x}, t) \cdot d \bm{x} \ = \ \frac{1}{2} \,\bar{\Phi} (t_f), 
\end{equation}
where we have introduced the following auxilliary function
\begin{equation}
 \bar{\Phi} (t_f) \ \equiv \ \frac{1}{t_f} \,\int_0^{t_f} \,\Phi (t) \, d t.
\end{equation}
Exactly in the same manner, we can show that
\begin{equation}
 \int_{C_2} \,\bm{A} (\bm{x}, t) \cdot d \bm{x} \ = \ \frac{- \,\omega_0}{2 \,\pi} \,
 \int_0^{t_f} \Phi (t) \, d t \ = \ - \,\frac{1}{2} \,\bar{\Phi} (t_f) 
\end{equation}
Combining the two line integrals, we are led to the following answer for
the time-dependent AB-phase shift
\begin{equation}
 \phi_{AB} \ = \ e \,\oint_{C_1 - C_2} \,\bm{A} (\bm{x}, t) \cdot d \bm{x}
 \ = \ e \,\bar{\Phi} (t_f) \ = \ e \,\,\,\frac{1}{t_f} \,\int_0^{t_f} \,\Phi (t) \,d t.
 \label{Eq:TD_AB-phase}
\end{equation}
We recall that the answer above is obtained under the condition that
the effect of the Lorentz force due to the induced electric field can approximately
be neglected. Remarkably, in a recent paper, Gao showed that the same answer 
as above can be obtained even without neglecting the effect of the 
induced electric field \cite{Gao2025}.
This is a highly non-trivial observation in view of the fact that the electron on the 
path $C_1$ and that on the path $C_2$ receives the Lorentz forces in the 
opposite directions relative to their moving directions.

\vspace{2mm}
In the following, we investigate the role of Lorentz force on the
electron following Gao's elegant formulation but with some additional care.
We first recall the fact that, since the electron can be thought of as moving 
in the $xy$-plane, the orbital angular momentum (OAM) of the electron has
only the $z$ component, which we simply denote as $L$.
Then, the change of the OAM $L$ under the presence of time-varying 
electromagnetic field is in general controlled by the following equation of 
motion \cite{Peshkin1981, OP1985, PT1989, WKZZ2018} : 
\begin{equation}
 \frac{d}{d t} \,L \ = \ - \,e \,\left[ \,\bm{x} \times 
 \left( \bm{E} + \bm{v} \times \bm{B} \right) \,\right]_z .
\end{equation}
Note that the r.h.s. of the above equation is nothing but the torque by the Lorentz 
force acting on the electron.  
With the use of the cylindrical coordinate $\bm{x} = (\rho, \phi, z)$, the above
equation can alternatively be expressed as
\begin{equation}
 \frac{d}{d t} \,L \ = \ - \,e \,\rho \, E_\phi \ + \ e \,\rho \,v_\rho \,B_z, 
\end{equation}
where $E_\phi$ and $B_z$ represent the azimuthal component of
the electric field and the $z$-component of the magnetic field, respectively,
while $v_\rho$ does the radial component of the electron's velocity.
For simplicity, we again assume that the two electron beams emitted from
the point $P$ move on the two circles $C_1$ and $C_2$ with the radius
$\rho \,(\,> R)$. Using the fact that, outside the solenoid, the electric and
magnetic fields are respectively given by
\begin{equation}
 E_\phi (\bm{x}, t) \ = \ - \frac{1}{2 \,\pi \,\rho} \,\frac{d}{d t} \,\Phi (t), \ \ \ 
 B_z (\bm{x}, t) \ = \ 0 \ \ \ \mbox{for} \ \ \ \rho > R,
\end{equation}
we are led to the relation
\begin{equation}
 \frac{d}{d t} \,L (t) \ = \ \frac{e}{2 \,\pi} \,\frac{d}{d t} \,\Phi (t).
\end{equation}
On the other hand, by using the relations $x = \rho \,\cos \phi$ and
$y = \rho \,\sin \phi$, the OAM $L$ can be expressed as
\begin{equation}
 L \ \equiv \ (\bm{x} \times \bm{p})_z \ = \ m \,( x \,\dot{y} - y \,\dot{x})
 \ = \ m \,\rho \,\dot{\phi},
\end{equation}
where $m$ denotes the mass of the electron.
Since the time derivative of the azimuthal angle $\phi$ is just
the angular velocity $\omega$ of the electron, i.e. $\omega (t) = \frac{d}{d t} \phi (t)$,
it follows the relation : 
\begin{equation}
 L (t) \ = \ m \,\rho \,\omega (t).
\end{equation}
Combining it with the equation of motion for $L (t)$, we are thus led to the 
equation \cite{Gao2025}
\begin{equation}
 \frac{d}{d t} \,\omega (t) \ = \ \frac{e}{2 \,\pi \,m \,\rho} \,\frac{d}{d t} \,
 \Phi (t).
\end{equation}
(Take care that the quantity $\Phi (t)$ on the r.h.s. represents the time-dependent
total magnetic flux inside the solenoid not the azimuthal angle $\phi$.)

\vspace{2mm}
Now Gao's argument goes as follows \cite{Gao2025}. 
The differential equation for $\omega (t)$
can easily be solved. On the semicircle $C_1$, the general solution is given as
\begin{equation}
 \omega_1 (t) \ - \ \omega_1 (0) \ = \ \frac{e}{2 \,\pi \, m \,\rho} \,
 \left[ \Phi (t) \ - \ \Phi (0) \right] .
\end{equation}
Similarly, the general solution on $C_2$ is given as
\begin{equation}
 \omega_2 (t) \ - \ \omega_2 (0) \ = \ \frac{e}{2 \,\pi \, m \,\rho} \,
 \left[ \Phi (t) \ - \ \Phi (0) \right] .
\end{equation}
Here, we set $t_i = 0$ as the initial time when the two electron beams are emitted from 
the point $P$, while we use $t_f = T$ as the time when the two electron beams
encounter again on the circle.
For easier comparison with the static magnetic flux case, we may take
\begin{equation}
 \omega_1 (0) \ = \ \omega_0, \ \ \ \omega_2 (0) \ = \ - \,\omega_0.
\end{equation}
This gives
\begin{eqnarray}
 \omega_1 (t) &=& + \,\omega_0 \ + \ \frac{e}{2 \,\pi \,m \,\rho} \,
 \left[ \Phi (t) \ - \ \Phi (0) \right]  \ \ \ \mbox{on} \ \ C_1, \\
 \omega_2 (t) &=& - \,\omega_0 \ - \ \frac{e}{2 \,\pi \,m \,\rho} \,
 \left[ \Phi (t) \ - \ \Phi (0) \right]  \ \ \ \mbox{on} \ \ C_2.
\end{eqnarray}
Remembering the expression $\bm{A} (\bm{x}, t) = \frac{\Phi (t)}{2 \,\pi \,\rho} \,
\bm{e}_\phi$ together with the relation $\bm{e}_\phi \cdot d \bm{x} = \rho \,d \phi
= \rho \,\omega (t) \,d t$, we therefore get
\begin{eqnarray}
 \int_{C_1} \,\bm{A} (\bm{x}, t) \cdot d \bm{x} &=& \int_0^T \, d t \,
 \frac{\Phi (t)}{2 \,\pi \,\rho} \,\, \rho \,\omega_1 (t) \ = \
 \frac{1}{2 \,\pi} \,\int_0^T \,d t \,\,\omega_1 (t) \,\Phi (t) \nonumber \\
 &=& 
 \frac{1}{2 \,\pi} \,\int_0^T \, d t \,\left\{ \,+ \,\omega_0 \ + \ \frac{e}{2 \,\pi \,m \,\rho} \,
 \left[ \Phi (t) \ - \ \Phi (0) \right] \right\} \,\Phi (t) .
\end{eqnarray}
Similarly, we obtain
\begin{eqnarray}
 \int_{C_2} \,\bm{A} (\bm{x}, t) \cdot d \bm{x}
 \ = \  
 \frac{1}{2 \,\pi} \,\int_0^T \,d t \,
 \left\{ \,- \,\omega_0 \ + \ \frac{e}{2 \,\pi \,m \,\rho} \,
 \left[ \Phi (t) \ - \ \Phi (0) \right] \right\} \,\Phi (t) .
\end{eqnarray}
Taking the difference between the above two line integrals, we therefore find that
\begin{equation}
 \oint_{C_1 - C_2} \,\bm{A} (\bm{x}, t) \cdot d \bm{x} \ = \ 
 \frac{1}{2 \,\pi} \,\int_0^T \,d t \,\,2 \,\omega_0 \,\Phi (t) \ = \ 
 \frac{1}{T} \,\int_0^T \,\Phi (t) \,d t,
\end{equation}
where use has been made of the relation $\omega_0 \,T = \pi$.
The desired AB-phase shift is then expressed as
\begin{equation}
 \phi_{AB} \ = \ e \,\oint_{C_1 - C_2} \,\bm{A} (\bm{x}, t) \cdot d \bm{x}
 \ = \ e \,\ \frac{1}{T} \,\int_0^T \,\Phi (t) \,d t.
\end{equation}
Using the time variable $t_f$ instead of $T$, this answer precisely coincide
with Eq.(\ref{Eq:TD_AB-phase}), which was obtained in \cite{Waka2025} 
by neglecting the effect of induced electric field. 
That the same answer is obtained even without neglecting
the effect of induced electric is a highly-nontrivial observation in view of
the fact that the electron on the path $C_1$ and that on the path $C_2$ 
feel Lorentz forces in the opposite directions relative to their moving directions.
At this point, the following question naturally arises. 
In the analysis by Gao \cite{Gao2025}, the time $t = T$ is simply stated as 
representing the time when the two beams ejected from the point $P$ overlap again
and re-interface. However, where on the circle these two beams re-encounter
is not clearly stated and left ambiguous. 
In the following, we shall clear up this mismatch.

\vspace{2mm}
What plays an important role here are the following two identities : 
\begin{equation}
 \phi_f \ = \ \int_{C_1} \,d \phi \ = \ \int_0^T \,\omega_1 (t) \,d t
 \ = \ \int_0^T \,\left\{ + \,\omega_0 \ + \ \frac{e}{2 \,\pi \,m \,\rho} \,
 \left[ \Phi (t) - \Phi (0) \right] \right\} \,d t,
\end{equation}
and
\begin{equation}
 \phi_f - 2 \,\pi \ = \ \int_{C_2} \,d \phi \ = \ \int_0^T \,\omega_2 (t) \,d t
 \ = \ \int_0^T \,\left\{ - \,\omega_0 \ + \ \frac{e}{2 \,\pi \,m \,\rho} \,
 \left[ \Phi (t) - \Phi (0) \right] \right\} \,d t.
\end{equation}
Taking the difference between the two relations, we get
\begin{equation}
 2 \,\pi \ = \ 2 \,\omega_0 \,\int_0^T \,d t \ = \ 2 \,\omega_0 \,T,
\end{equation}
which just reconfirms the relation $T = \pi / \omega_0$.
On the other hand, taking the sum of the two equations, we obtain
\begin{equation}
 \phi_f \ = \ \pi \ + \ \frac{e}{2 \,\pi \,m \,\rho} \,\int_0^T \,
 \left[ \Phi (t) \ - \ \Phi (0) \right] \,d t .
\end{equation}
This dictates that the re-encountering place $Q$ of the two electron beams
is not just the opposite side of $P$ on the circle but it is specified by
the azimuthal angle $\phi_f$ given by the above equation. Note that this angle
is dependent on the arrival time $T$ of the electron beams as well as on
several other parameters, i.e. $e$, $m$ and $\rho$. 
(This raises some additional questions. We shall later come back to this issue
in the next section.)
Here, to get an overview of what is happening, we think it instructive to 
consider one specific example for the time-dependent magnetic flux $\Phi (t)$.

\begin{figure}[ht]
\begin{center}
\includegraphics[width=10.0cm]{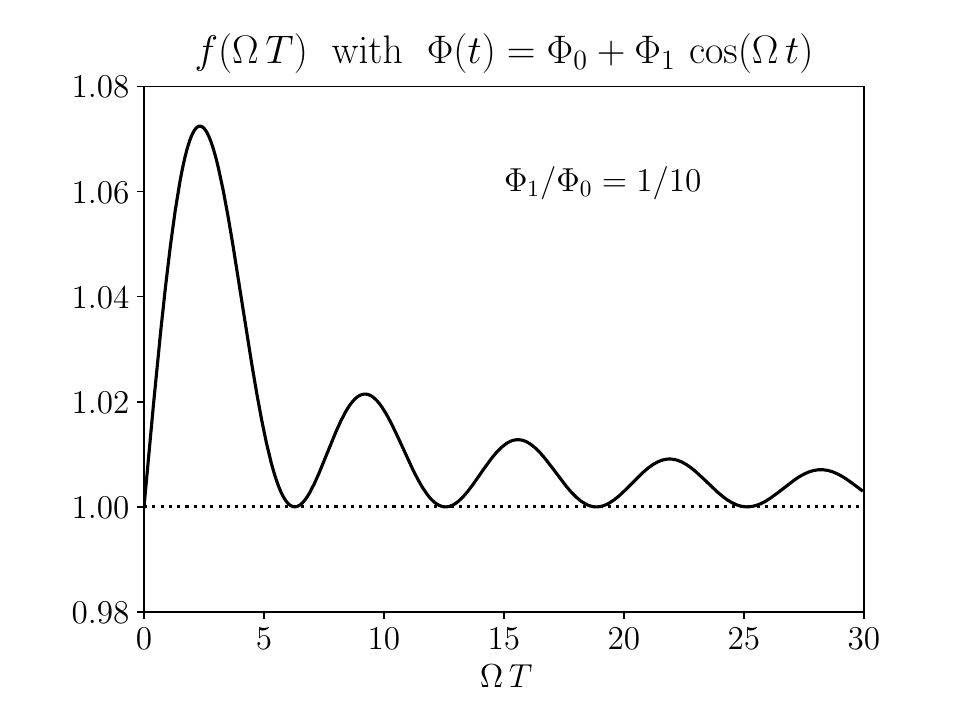}
\caption{The auxiliary function $f (\Omega \,T)$ defined in Eq.(\ref{Eq:f}) corresponding
to the time-dependent magnetic flux $\Phi (t) = \Phi_0 + \Phi_1 \,\cos \Omega t$.
The figure corresponds to the choice $\Phi_1 / \Phi_0 = 1 / 10$.}
\label{Fig:TD_AB-phase}
\end{center}
\end{figure}

\vspace{2mm}
As an illustrative example, let us consider the setting in which $\Phi (t)$ is given 
as a sum of the time-independent part and the sinusoidally oscillating
part as
\begin{equation}
 \Phi (t) \ = \ \Phi_0 \ + \ \Phi_1 \,\sin \Omega \,t,
\end{equation}
where $\Phi_0$ and $\Phi_1$ are some constants.
In this case, the time-dependent AB-phase shift is given as
\begin{equation}
 \phi_{AB} \ = \ e \,\,\frac{1}{T} \,\int_0^T \,\Phi (t) \,d t \ = \ 
 e \,\left[\, \Phi_0 \ - \ \Phi_1 \,
 \left( \frac{\cos \Omega \,T}{\Omega \,T} \ - \ \frac{1}{\Omega \,T}
 \right) \right] \ = \ e \,\Phi_0 \,f (\Omega \,T),
\end{equation}
with
\begin{equation}
 f (\Omega \,T) \ = \ 1 \ + \ \frac{\Phi_1}{\Phi_0} \,
 \left( \frac{\cos \Omega \,T}{\Omega \,T} \ - \ \frac{1}{\Omega \,T}
 \right) . \label{Eq:f}
\end{equation}
To see the dependence of $\phi_{AB}$ on the arrival time $T$, we show
in Fig.\ref{Fig:TD_AB-phase} the function $f (\Omega \,T)$ defined above for 
the sample choice $\Phi_1 / \Phi_0 = 1 / 10$. One sees that the function 
$f (\Omega \,T)$ clearly shows an oscillatory behavior as a function of 
$\Omega \,T$, which obviously indicates the existence of the time-dependent 
AB-effect.
 
\vspace{2mm}
One may also be interested in where on the circle the two electron beams 
re-encounter. By using the relation
\begin{equation}
 \int_0^T \,\left[ \Phi (t) \ - \ \Phi (0) \right] \, d t \ = \
 - \,\Phi_1 \,\left( \frac{\cos \Omega T}{\Omega} \ - \ \frac{1}{\Omega}
 \right), 
\end{equation}
we find that
\begin{equation}
 \phi_f \ = \ \pi \ - \ \frac{e \,\Phi_1}{2 \,\pi \,m \,\rho} \,T \,
 \left( \frac{\cos \Omega T}{\Omega \,T} \ - \ \frac{1}{\Omega \,T}
 \right), \label{Eq:phi_f}
\end{equation}
which clearly shows that $\phi_f$ is in general different from $\pi$.
It would be instructive to inspect some limiting cases.
First, in the limit $\Omega \rightarrow 0$, one can readily confirm the 
following behavior
\begin{equation}
 \phi_{AB} \ \rightarrow \ e \,\Phi_0 \ \ \mbox{and} \ \   
 \phi_f \ \rightarrow \ \pi. 
\end{equation}
This is only natural, since this limit corresponds to the case of
familiar time-independent AB-effect. 

\vspace{2mm}
On the other hand, in the opposite limit $\Omega \rightarrow \infty$, which
corresponds to the case where the oscillation period is extremely short,
one finds that $\phi_{AB}$ and $\phi_f$  also approaches 
$e \,\Phi_0$ and $\pi$, i.e.
\begin{equation}
 \phi_{AB} \ \rightarrow \ e \,\Phi_0 \ \ \ \mbox{and} \ \ \  
 \phi_f \ \rightarrow \ \pi. 
\end{equation}
although, in this case,  both approach their limiting values while oscillating 
as a function of $T$.

\section{Questions still remain and outlook}
\label{Section:s3}

From the theoretical analyses so far, we have definitely confirmed the likely 
existence of the phenomenon called the time-dependent Aharonov-Bohm effect.  
We could also understand the role of the induced electric field generated
by the time-varying magnetic flux inside the solenoid at least partially.
Still, we must say that the present analyses of the time-dependent AB-effect are 
far from complete and leave behind a difficult question. 
The nature of the problem may be understood
if one compares the prediction for the AB-phase shift $\phi_{AB}$ and
that for the angle $\phi_f$, both of which are anyhow obtained after taking
account of the induced electric field. First, pay attention to the fact
that the expression for the time-dependent AB-phase $\phi_{AB}$
given by Eq.(\ref{Eq:phi_f}) depends only on the time-dependent magnetic flux $\Phi (t)$
inside the solenoid aside from the arrival time $T$ of the electron beams.
This resemblance with the familiar time-independent AB-phase shift 
implies the topological nature of the time-dependent AB-phase shift as
indicated in the Appendix of \cite{Waka2025}.  
On the other hand, the expression for the azimuthal angle $\phi_f$, which specifies
the re-encoutering place of the two electron beams on the circle, depends not only
on the magnetic flux $\Phi (t)$ but also on several other quantities like the mass 
$m$ and the charge $e$ of the electron and the chosen radius $\rho$ of the traveling
circle of the electron beams. In particular, the dependence on the radius $\rho$
definitely indicates that the angle $\phi_f$ is not a quantity which can be given 
any topological significance and that the prediction (\ref{Eq:phi_f}) might be meaningful
only within specific choice of the semi-circular paths $C_1$ and $C_2$.  

\begin{figure}[ht]
\begin{center}
\includegraphics[width=10.0cm]{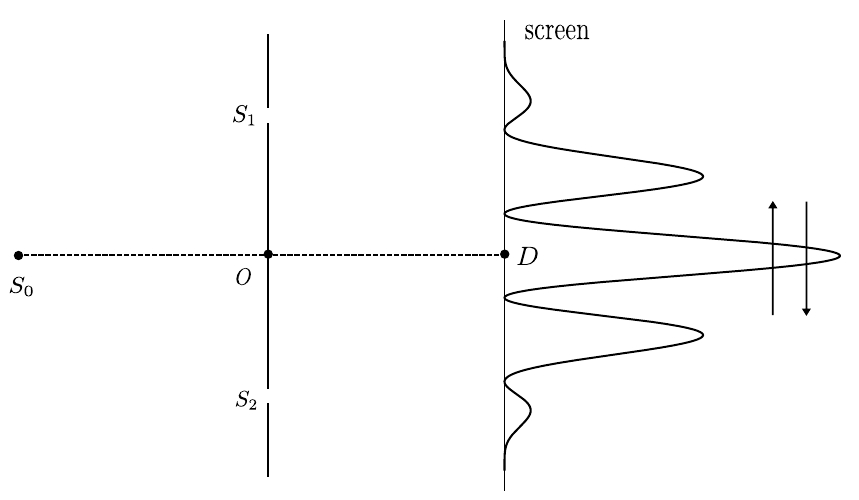}
\caption{The picture schematically showing the oscillatory behavior of the peak
position of the interference pattern on the screen as the arrival time $T$ or 
equivalently the velocity $v$ of the electron beams changes.}
\label{Fig:AB-phase_interference}
\end{center}
\end{figure}

Incidentally, the reason why we have considered the circular paths of the 
electron should be obvious.  
It is because this choice respects the axial symmetry of the problem and it
is most convenient for handling the effect of the induced electric field.
(Remember that the Lorentz force acting on the electron due to the induced
electric field acts on the circumferential direction.)
In the theoretical analysis of the AB-effect,  there is still another popular setting
for handling the problem.
It is the familiar interference measurement using a double-slit screen and another 
screen behind it for observing the interference fringes. In this consideration, the electron 
beams are treated as free waves which are emitted from an ejector $S_0$ and travel
in straight lines through the two slits on the intermediate screen and finally reach 
the interference fringe observation screen.
Some years ago, Choudhury and Mahajan carried out what they call a
direct calculation of time-dependent AB-phase shift  in the schematic setting
of the double-slit measurement \cite{CM2019}. 
The form of the sinusoidally oscillating magnetic flux is not the same as that 
considered in the present paper.
Besides, their expression for the time-dependent AB-phase shift was 
obtained by using several approximations and it takes a slightly complicated 
form as compared with our analytic expression.
Nonetheless, their theoretical prediction for the time-dependent AB-phase shift
also shows oscillatory behavior as a function of the arrival time $T$,
which gives an additional support to the very existence of the time-dependent 
AB-effect. Although it is not discussed in their paper, we think it is interesting
to pursue the possibilities as described below.

\vspace{1mm}
To explain the idea more concretely, suppose that the point-like slits 
$S_1$ and $S_2$ are located symmetrically on the $y$-axis passing through the origin $O$
as illustrated in Fig.\ref{Fig:AB-phase_interference}.
The point $D$ on the observation screen is specified by the line $S_0OD$ from the 
source $S_0$ to the point $D$ along the $x$-axis with $y = 0$.
In the case of standard time-independent magnetic flux, it is obvious that the point $D$
on the screen is the peak position of interference pattern on the observation screen.
As can be naturally imagined from our present analysis, if the magnetic flux is sinusoidally 
oscillating, the peak position of interference pattern is expected to oscillate around the 
point $D$ as a function of the arrival time $T$ of the electron beam. 
The arrival time $T$ of the electron
beams is thought to inversely proportional to the velocity $v$ of the electron beam.
It should be contrasted to the fact that the standard AB-effect is independent
of the electron velocity and this fact is sometimes called the {\it dispersionless nature} of
the AB-effect \cite{Zellinger1986, Peshkin1999, CBB2007}. 
In other words, the dispersionless nature is not the property 
of the time-dependent AB-effect. From the consideration above, we rather anticipate 
that the peak position of interference on the screen would oscillate around the point 
$D$  as a function of the velocity $v$ of the electron beam. 
At the present stage, it is difficult to give a fully quantitative prediction
on the time variation of the peak position of interference pattern.
The reason is because now we do not have any satisfactory understanding 
about the electron path dependence or independence for the prediction of the 
interference peak position. Still, the change of the peak position of interference pattern 
as a function of the electron velocity is a physical phenomenon that is most likely to 
occur and it is obviously an experimentally measurable object. 
We anticipate that such a measurement would provide us with the simplest 
means for confirming the existence or non-existence of the long-unconfirmed 
time-dependent AB-effect.




\vspace{1mm}
\noindent
\section*{References}

\bibliography{TDABbibfile}

@article{ES1949,
 title = "{The refractive index in electron optics and the principles of dynamics}",
 journal = "Proc. Phys. Soc. London",
 volume = "B62",
 pages = "8",
 year = "1949",
 author = "W.~Ehrenberg and R. E.~Siday"
}

@article{AB1959,
 title = "{Significance of electromagnetic potentials in quantum theory}",
 journal = "Phys. Rev.",
 volume = "115",
 pages = "485",
 year = "1959",
 author = "Y.~Aharonov and D.~Bohm"
}

@article{Tonomura1986,
 title = "{Evidence for Aharonov-Bohm effect with magnetic field completely
 shielded from electron wave}",
 journal = "Phys. Rev. Lett.",
 volume = "56",
 pages = "792",
 year = "1986",
 author = "A.~Tonomura and N.~Osakabe and T.~Matsuda and T.~Kawasaki and
 J.~Endo and S.~Yano and H.~Yamada"
}

@article{Osakabe1986,
 title = "{Experimental confirmation of Aharonov-Bohm effect using a
 toroidal magnet field confined by a superconductor}",
 journal = "Phys. Rev.",
 volume = "A34",
 pages = "815",
 year = "21986",
 author = "N.~Osakabe and T.~Matsuda and T.~Kawasaki and J.~Endo and A.~Tonomura 
 and S.~Yano"
}

@article{Peshkin1981,
 title = "{The Aharonov-Bohm effect : Why it cannot be eliminated from quantum mechanics}",
 journal = "Phys. Rep.",
 volume = "80",
 pages = "375",
 year = "1981",
 author = "M.~Peshkin"
}

@article{OP1985,
 title = "{The quantum effects of electromagnetic fluxes}",
 journal = "Rev. Mod. Phys.",
 volume = "57",
 pages = "339",
 year = "1985",
 author = "S.~Olariu and I. I.~Popescu"
}

@article{PT1989,
 title = "{The Aharonov-Bohm effect}",
 journal = "Lecture Note in Physics",
 volume = "340",
 pages = "1",
 year = "1989",
 author = "M.~Peshkin and A.~Tonomura"
}

@article{WKZZ2018,
 title = "{The role of electron orbital angular momentum in the
 Aharonov-Bohm effect revisited}",
 journal = "Ann. Phys.",
 volume = "397",
 pages = "259",
 year = "2018",
 author = "M.~Wakamatsu and Y.~Kitadono and L.~Zou and P.-M. Zhang"
}

@article{Waka2025,
 title = "{On the time-dependent Aharonov-Bohm effect and the
 4-dimensional Stokes theorem}",
 journal = "Ann. Phys.",
 volume = "477",
 pages = "169984",
 year = "2025",
 author = "M.~Wakamatsu"
}

@article{SV2013,
 title = "{The covariant, time-dependent Aharonov-Bohm effect}",
 journal = "Phys. Lett.",
 volume = "B723",
 pages = "241",
 year = "2013",
 author = "D.~Singleton and E.C.~Vagenas"
}

@article{MS2014,
 title = "{Stokes' theorem, gauge symmetry and time-dependent Aharonov-Bohm effect}",
 journal = "J. Math. Phys. ",
 volume = "55",
 pages = "042101",
 year = "2014",
 author = "J.~Macdougall and D.~Singleton"
}

@article{AS2016,
 title = "{Time dependent electromagnetic fields and 4-dimensional Stokes' theorem}",
 journal = "Am. J. Phys.",
 volume = "84",
 pages = "848",
 year = "2016",
 author = "R.~Andosca and D.~Singleton"
}

@article{LYGC1992,
 title = "{Analysis of Aharonov-Bohm effect due to time-dependent vector potentials}",
 journal = "Phys. Rev. ",
 volume = "A45",
 pages = "4319",
 year = "1992",
 author = "B.~Lee and E.~Yin and T.K.~Gustafson and R.~Chiao"
}

@article{ADC2000,
 title = "{Magnetic Aharonov-Bohm Effect under Time-Dependent Vector Potential}",
 journal = "Technical Physics Letters",
 volume = "26",
 pages = "392",
 year = "2000",
 author = "A.N.~Ageev and S.Yu.~Davydov and A.G.~Chirkov"
}

@article{GNS2011,
 title = "{Homotopy and Path Integrals in the Time-dependent Aharonov-Bohm Effect}",
 journal = "Found. Phys. ",
 volume = "41",
 pages = "1462",
 year = "2011",
 author = "B.~Gaveau and A.M.~Nounou and L.S.~Shulmann"
}

@article{JZWLD2017,
 title = "{On the time-dependent Aharonov-Bohm effect}",
 journal = "Phys. Lett.",
 volume = "B774",
 pages = "87",
 year = "2017",
 author = "J.~Jing and Y.-F.~Zhang and K.~Wang and Z.-W.~Long and S.-H.~Dong"
}

@article{CM2019,
 title = "{Direct calculation of time varying Aharonov-Bohm effect}",
 journal = "Phys. Lett. ",
 volume = "A384",
 pages = "2467",
 year = "2019",
 author = "S. R.~Choudhury and S.~Mahajan"
}

@misc{Gao2025,
 title = "{Generalized Aharonov-Bohm effect : Its derivation, theoretical implications
 and experimental tests}",
 author = "S.~Gao",
 year = "2025",
 howpublished = "\url{https://philsci-archive.pitt.edu/24726/}"
}

@inProceedings{Zellinger1986,
 title = "{Generalized Aharonov-Bohm Experiments with Neutrons}",
 booktitle = "Fundamental Aspects of Quantum Theory, Como 1985;",
 journal = "Physica B",
 volume = "269",
 pages = "311-318",
 year = "1986",
 author = "A.~Zellinger"
}

@article{Peshkin1999,
 title = "{Force Free Interactions and Nondispersive Phase Shift in Interferometry}",
 journal = "Found. Phys.",
 volume = "29",
 pages = "481",
 year = "1999",
 author = "M.~Peshkin"
}

@article{CBB2007,
 title = "{Macroscopic Test of the Aharonov-Bohm Effect}",
 journal = "Phys. Rev. Lett.",
 volume = "99",
 pages = "210401",
 year = "2007",
 author = "A.~Caprez and B.~Barwick and H.~Batelaan"
}
\bibliographystyle{unsrt}





\end{document}